\documentstyle[preprint,epsfig,aps]{revtex}

\def\Journal#1#2#3#4{{#1} {\bf #2}, #3 (#4)}

\def\CJF{\em Czech. J. Phys}
\def\IZV{\em Izv. AN SSSR}
\def\JINR{\em JINR Communications}
\def\PPNP{\em Prog. Part. Nucl. Phys.}
\def\PTP{\em Prog. Theor. Phys. (Supp.) }

\def\NPA{{\em Nucl. Phys.} A}
\def\PLB{{\em Phys. Lett.}  B}
\def\PRL{\em Phys. Rev. Lett.}

\def\PRC{{\em Phys. Rev.} C}
\def\PRP{\em Phys. Rep.} 

\def\ZPA{{\em Z. Phys.} A}

\def\RPP{\em Rep.Prog. Phys.} 
\def\RMP{\em Rev. Mod. Phys.} 
\def\FIZB{{\em FIZIKA} B} 
\def\be{\begin{equation}}
\def\ee{\end{equation}}
\def\bea{\begin{eqnarray}}
\def\eea{\end{eqnarray}}
\begin{document}
\draft                                                            

\title{Two-Neutrino Double Beta Decay: Critical Analysis}

\author{ F. \v SIMKOVIC }

\address{Department of Nuclear Physics, Comenius University \\
SK-84215 Bratislava, Slovakia}

\author{ G. PANTIS }

\address{Department of Physics, University of Ioannina,\\ 
GR-45100 Ioannina, Greece}

\author{ AMAND FAESSLER }

\address{Institute f\"ur Theoretische Physik der Universit\"at 
T\"ubingen\\ 
Auf der Morgenstelle 14, D-72076 T\"ubingen, Germany}

\maketitle

\begin{abstract}
We have performed a critical analysis of different approximation
schemes for the calculation of two-neutrino double beta decay
($2\beta 2\nu$-decay) matrix elements. For that purpose the time
integral representation of the  $2\beta 2\nu$-decay matrix element
has been used. We have shown that within the  single-particle
approximation of the nuclear Hamiltonian the $2\beta 2\nu$-decay
matrix element is equal to zero because of the mutual cancelation 
of the direct and cross terms. The quasiboson approximation (QBA)
and renormalized QBA (RQBA) schemes imply for the 
$2\beta 2\nu$-decay transition operator to be a constant, if one
requires the equivalence of initial and final Quasiparticle Random
Phase Approximation (QRPA) and renormalized QRPA (RQRPA) Hamiltonians.
It means that $2\beta 2\nu$-decay is a higher order process in the
boson expansion of the nuclear Hamiltonian and its higher order boson
approximations are important. The equivalence of the initial and 
final QRPA  and RQRPA Hamiltonians has been discussed within the 
QBA and RQBA, respectively. 
We have found that the mismatching of both Hamiltonians
is getting worse with increasing strength of particle-particle 
interaction especially in the case of QRPA Hamiltonians. It is 
supposed to be one of the reasons of the extreme sensitivity of 
the $2\beta 2\nu$-decay matrix element to the residual interaction
appearing in explicit calculations involving the intermediate nucleus.
Further, the Operator Expansion Method (OEM) has been reconsidered and new 
$2\beta 2\nu$-decay transition operators have been rederived in
a consistent way. The validity of the  OEM approximation has been
discussed in respect to the other approximation schemes. 
The OEM combined with QRPA or RQRPA ground state
wave functions reflects sensitively the instabilities 
incorporated in the considered ground states. 
Therefore, the predicting power of the OEM should be studied 
with help of the other ground state wave functions,
e.g. shell model ones. 
\end{abstract}
\pacs{23.40.Hc}

\section{ Introduction}

The two-neutrino double beta decay ($2\beta 2\nu$-decay) is a second 
order process of the weak interaction, which is allowed by the 
Standard model\cite{1}. 
In $2\beta 2\nu$-decay the nucleus (A,Z) undergoes the 
transition to nucleus (A,Z+2) with emission of two electrons and 
two antineutrinos. This rare process is already well established
experimentally for a couple of isotopes. The inverse half-live of 
$2\beta 2\nu$-decay is free of unknown parameters 
on the particle physics side and is expressed as a product 
of  a phase-space factor and the relevant $2\beta 2\nu$-decay
nuclear matrix element. Since the phase-space factor can be calculated 
with the desired accuracy, the experimental studies of $2\beta 2\nu$-decay 
give us directly the value of the $2\beta 2\nu$-decay 
nuclear matrix element. In this way $2\beta 2\nu$-decay offers a
sensitive test of nuclear structure calculations.
The calculation of the  $2\nu\beta\beta$-decay 
nuclear transition continues to be challenging 
in view of the smallness of the predicted nuclear matrix elements 
and the fact that the mechanism 
which is leading to the suppression of these matrix elements is still not 
completely understood. 

The proton-neutron Quasiparticle Random Phase Approximation (pn- QRPA) 
has been the most frequently used nuclear structure method for evaluating 
$2\beta 2\nu$-decay rates because of the remarkable success achieved
in revealing the suppression mechanism of $2\beta 2\nu$-decay matrix 
elements\cite{2}$^{-}$\cite{4}. 
However, the extreme sensitivity of the $2\beta 2\nu$-decay matrix 
elements on the pn $1^+$ particle-particle matrix element 
as well as the collapse of the QRPA solution 
in the physically acceptable region of the particle-particle 
strength of the nuclear Hamiltonian renters it difficult to make definite
rate predictions. 

Some attempts have been done to overcome the above drawbacks, e.g. by
including higher order RPA corrections\cite{5}, 
particle number projection\cite{6,7} and 
proton-neutron pairing\cite{8} 
in the theory. However, none of these modifications 
of the QRPA prevents the collapse and inhibit the nuclear matrix element 
to cross zero close to the physical value of the particle-particle force.
Recently, Toivanen and Suhonen have proposed a 
proton-neutron renormalized QRPA (pn-RQRPA)\cite{9}, which 
goes beyond the QRPA and takes into account the 
Pauli exclusion principle in an approximate way. It has been shown 
that the above  phenomena could be connected with the limitation of the
QRPA approach - the quasiboson approximation, 
which violates the Pauli exclusion principle. The renormalized 
quasiboson approximation (QBA), on which the pn-RQRPA is based, 
inhibits the collapse of the pn-RQRPA solution for a physical value
of the particle-particle interaction strength.
In addition, the calculated
$2\beta 2\nu$-decay nuclear matrix elements via pn-RQRPA have been found 
significantly less sensitive to the particle-particle force
within its physical values in respect to those obtained by the 
pn-QRPA\cite{9}$^{-}$\cite{12}. 
This behavior  has been confirmed also within the renormalized QRPA with 
proton-neutron pairing (full-RQRPA)\cite{10}.  

In spite of the advantages, of the renormalized QRPA over the QRPA,
the RQRPA can not be considered  as the ultimate solution for the calculation
of the $2\beta 2\nu$-decay process. Several 
shortcomings still
plague the RQRPA: i) The Ikeda sum rule is violated\cite{12,13}.
ii) There are  two sets of the intermediate nuclear states 
in the calculation generated respectively from initial and final nuclei, 
which do not coincide with each other. iii) The Pauli exclusion 
principle is taken into account only in an approximate way. 
All these leaps of faith 
of the RQRPA approach have common origin. It is the particle number 
non-conservation. 
The effect of these shortcomings on the $2\beta 2\nu$-decay amplitude 
is well understood. 

A longstanding problem of large discrepancies between the values 
of the predicted and calculated $2\beta 2\nu$-decay matrix elements
has led to a development of alternative methods, e.g. one of them is the
Operator Expansion Method (OEM)\cite{14}$^{-}$\cite{23}. 
The OEM tries to avoid the necessity
of evaluating the sum over the intermediate nuclear states.
The price paid for it is that one has now to deal with more than
two-body operators and commutators involving kinetic energy terms in 
the commutator expansion of the $2\beta 2\nu$-decay transition
operator. Recently, this method has been reconsidered\cite{23} 
and it has been 
shown that the previous derivation\cite{14}$^{-}$\cite{21} 
of the OEM-potential was not consistent.
The OEM-calculation with a consistent OEM-potential combined with the
pn-RQRPA ground state way functions (OEM+RQRPA) has exhibited a large 
sensitivity  on the strength of the particle-particle 
force within its physical values\cite{23}. There is a speculation that 
the approximations of the pn-RQRPA method are responsible  for this behavior.
Therefore,  detail  OEM+RQRPA calculations are expected to be helpful for 
solving the  problem of stability of the
$2\beta 2\nu$-decay matrix element  in respect to model parameters.

Till now no consistent many body approach is available for the
calculation of the many-body Green function governing
the $2\beta 2\nu$-decay process, because of the computational 
complexity of the problem. Therefore we can not avoid  the introduction
of different approximation schemes in the evaluation of the nuclear 
matrix elements. Nevertheless, we can try to understand the limitations 
of the different approximation schemes.  The aim of this work is to perform 
a critical analysis of the QBA and renormalized QBA schemes by using
the time integral representation of the  $2\beta 2\nu$-decay nuclear
matrix element and to discuss the validity of the QRPA, RQRPA and
OEM+RQRPA calculations.

\section{ $2\beta 2\nu$-decay nuclear matrix element.}

If the two-nucleon mechanism for the $2\beta 2\nu$-decay process is considered,
then for the matrix element of this process  we have
\begin{eqnarray}
<{f}|S^{{(2)}}|{i}>
\hspace{3cm}\nonumber \\
=\frac{(-{i})^{{2}}}{2}
{\left(\frac{G_{{F}}}{~\sqrt{2}}\right)}^{{2}}
N_{{{p}}_{{1}}} N_{{{p}}_{{2}}} 
N_{{{k}}_{{1}}} N_{{{k}}_{{2}}}
J_{\mu\nu}(p_{{1}},p_{{2}},k_{{1}},k_{{2}}) 
\nonumber \\
\times\bar{u}(p_{{1}})
\gamma_\mu(1+\gamma_{{5}})u(-k_{{1}})
\bar{u}(p_{{2}})
\gamma_\nu(1+\gamma_{{5}})u(-k_{{2}})
\nonumber \\
-(p_{{1}}\leftrightarrow p_{{2}}) - 
(k_{{1}}\leftrightarrow k_{{2}}) + 
(p_{{1}}\leftrightarrow p_{{2}})
(k_{{1}}\leftrightarrow k_{{2}}),
\label{eq.1} 
\end{eqnarray}
where
\begin{eqnarray}
J_{\mu\nu}(p_{{1}},p_{{2}},k_{{1}},k_{{2}})
=\int {{e}}^{{-i}{{(p}}_{{1}}
{{+k}}_{{1}}{{)x}}_{{1}}}
{{e}}^{{-i}{{(p}}_{{2}}
{{+k}}_{{2}}{{)x}}_{{2}}}\nonumber \\
{ _{{out}}<}p_{{f}}|T(J_\mu(x_{{1}})
J_\nu(x_{{2}}))|p_{{i}}>_{{in}} {d}x_{{1}} 
{d}x_{{2}}.
\label{eq.2}
\end{eqnarray}
Here, $N_{{p}}=(1/(2\pi)^{{3/2}})(1/(2p_0)^{{1/2}})$, 
$p_{{1}}$ and $p_{{2}}$ ($k_{{1}}$ and
$k_{{2}}$) are four-momenta of electrons (antineutrinos), 
$p_{{i}}$ and $p_{{f}}$ are four-momenta of the initial and final nucleus.
$J_\mu(x)$ is the weak charged nuclear hadron current in the Heisenberg 
representation\cite{24,25}. 

The matrix element in Eq.\ (\ref{eq.1}) contains the contributions 
from two subsequent nuclear beta decay processes and 
$2\beta 2\nu$-decay\cite{26}. They could be separated, if we write the
 T-product 
of the two hadron currents as follows:
\begin{eqnarray}
T(J_\mu(x_{{1}}) J_\nu(x_{{2}})) 
=J_\mu(x_{{1}}) J_\nu(x_{{2}}) + 
\Theta(x_{{20}} - x_{{10}})
[J_\nu(x_{{2}}),J_\mu(x_{{1}})].
\label{eq.3}
\end{eqnarray}
The first term in the r.h.s. of Eq.\ (\ref{eq.3}) is associated with
two subsequent nuclear beta decay processes, which are 
energetically forbidden for the most of 
$2\beta 2\nu$-decay isotopes. The second term 
corresponds to $2\beta 2\nu$-decay process. We see that  the
$2\beta 2\nu$-decay nuclear matrix element is given by the 
non-equal-time commutator of the two hadron currents. It will be shown 
later that this feature is crucial for our understanding of the 
approximation schemes of different nuclear models.

We further assume the following standard approximations:
i) The non-relativistic impulse approximation for the hadronic
current $J_\nu (0,\vec{y})$. ii) We keep only the contribution from the 
axial current. iii) Only the $s_{1/2}$
wave states of the emitted electrons are considered. iv)
Our interest will be restricted only to the most favored
$0^{{+}}_{{initial}} \rightarrow 0^{{+}}_{{final}}$ nuclear transition.  
Then we have,
\begin{eqnarray}
J^{2\beta2\nu}_{\mu\nu}(p_{{1}},
p_{{2}},k_{{1}},k_{{2}})=
-i2 M_{{GT}}\delta_{\mu {k}}
\delta_{\nu {k}} ~~~~\nonumber \\
\times 2\pi\delta(E_{{f}}-E_{{i}}+
p_{{10}}+k_{{10}}+p_{{20}}+k_{{20}})
,~k=1,2,3,
\label{eq.4}
\end{eqnarray}
where,
\begin{eqnarray}
M_{GT} = \frac{i}{2} \int_0^\infty 
(e^{it(p_{10}+k_{10}-\Delta)}+e^{it(p_{20}+k_{20}-\Delta)})
M_{AA}(t) dt,
\label{eq.5} 
\end{eqnarray}
with
\begin{eqnarray}
M_{AA}(t) = 
^{}_{f}<0^{{+}}|\frac{1}{2} 
[ A_{{k}}(t/2), A_{{k}}(-t/2) ] |0^{{+}}>_i,
~~~A_k (t) = e^{i H t} A_k (0) e^{- i H t}.
\label{eq.6} 
\end{eqnarray}
Here, $|0^{{+}}>_i$ and $|0^{{+}}>_f$ 
 are respectively the wave functions of the initial and  final  
nuclei with their corresponding energies $E_{{i}}$ and $E_{{f}}$. 
$\Delta$ denotes
the average energy $\Delta = (E_{{i}}-E_{{f}})/2$. 
$A_{{k}}(0)$ is the
Gamow-Teller transition operator $A_{{k}}(0)=\sum_{{i}} 
\tau^{{+}}_{{i}}(\vec{\sigma}_{{i}})_{{k}}$, 
k=1,2,3.

The time integral form of $M_{GT}$ in Eqs.\ (\ref{eq.5}) and (\ref{eq.6}) 
has been the starting point for the methods, which avoid
the explicit calculation of the intermediate nuclear states, e.g.
the Operator Expansion Method (OEM)\cite{15,16}$^{,}$\cite{22,23}
 and the S-matrix approach\cite{26}. 
The S-matrix approach requires the derivation of two-body operators from the 
corresponding exchange Feynmann 
diagrams and the calculation of the nuclear transition by using
a given initial and final nuclear wave functions. 

For an analytical study of the different approximation schemes,
it is useful to rewrite the transition
operator of the nuclear matrix element $M_{AA}(t)$  
in Eq.\ (\ref{eq.5}) into a infinite series 
of multiple commutators of the nuclear Hamiltonian H and the
Gamow Teller transition operator $A_k(0)$ with help of 
\begin{eqnarray}
A_k(t)
=\sum^{\infty}_{{n=0}}\frac{(it)^{{n}}}{n!}
\overbrace{[H[H...[H}^{{n}~{times}}
,A_k(0)]...]].
\label{eq.7}
\end{eqnarray}
If the multiple commutator is calculated without approximation
for a nuclear Hamiltonian consisting of one- and two-
body interactions, we obtain an
infinite sum of many-body operators. This difficulty 
may be avoided if some approximation schemes are introduced, e.g.
the QBA, the renormalized QBA or the approximation schemes of the 
Operator Expansion Method (OEM). We shall discuss this point in 
the next Section.

If we  integrate over the time variable in 
Eq.\ (\ref{eq.7})) using the
standard procedure of the adiabatic switch-off 
of the interaction as $t\to \infty$,
insert the complete set of the intermediate states $|1^+_n>$ 
with eigenenergies $E_{{n}}$
between the two axial currents in Eq.\ (\ref{eq.6}) and
assume that the nuclear states are eigenstates of the nuclear
Hamiltonian, we get the
well-known form of $M_{GT}$
\begin{eqnarray}
M_{{GT}}&=&\frac{1}{2}\sum_{{n}}
<0^{{+}}_{{f}}|A_{{k}}(0)|1^{{+}}_{{n}}>
<1^{{+}}_{{n}}|A_{{k}}(0)|0^{{+}}_{{i}}>\times \nonumber \\
&&\big(
\frac{1}{E_{{n}}-E_{{i}}+p_{10}+k_{10}} + 
\frac{1}{E_{{n}}-E_{{i}}+p_{20}+k_{20}} 
\big).
\label{eq.8}
\end{eqnarray}
After the usual approximation 
$p_{10}+k_{10} \approx p_{10}+k_{10} \approx \Delta$
the form of $M_{GT}$ in Eq.\ (\ref{eq.8}) 
is suitable for the calculation within the commonly used 
intermediate nucleus approaches (INA) to $2\beta 2\nu$-decay process
like QRPA, RQRPA  and shell model methods, which construct the spectrum 
of the intermediate nucleus by  diagonalization.

\section{ The approximation schemes}

\subsection{ Single particle Hamiltonian}

Let restrict our consideration  to a single particle 
Hamiltonian, which in the second quantization formalism takes
the form
\begin{equation}
{\hat{H}}^{{s.p.}}=\sum _{{p, m}_{{p}}}
e_{{p}}c^{{+}}_{{p,m}_{{p}}}
c^{}_{{p,m}_{{p}}}+
\sum _{{n, m}_{{n}}}e_{{n}}
c^{{+}}_{{n,m}_{{m}}}
c^{}_{{n,m}_{{n}}}.
\label{eq.9}
\end{equation}
Here, $c^{{+}}_{{p,m}_{{p}}}$ and 
$c^{{+}}_{{n,m}_{{n}}}$ 
($c^{}_{{p,m}_{{p}}}$ 
and $c^{}_{{n,m}_{{n}}}$) 
are creation (annihilation) operators
of proton and neutron, respectively and  
$e_{{p}}$ and $e_{{n}}$ are single
particle energies of proton and neutron states.
By using of the Eq.\ (\ref{eq.7}) and the anticommutation 
relation of particle operators for the time 
dependent axial current $\hat{A}_k(t)$ we obtain 
\begin{eqnarray}
{\hat{A}}^{s.p.}_k(t)
= \sum_{{p}, {m}_{{p}},
{n}, {m}_{{n}}}
{{e}}^{{i(e}_{{p}}
{-e}_{{n}}{)t}} 
<p,m_{{p}}|A_k(0)|n,m_{{n}}>
c^{{+}}_{{p,m}_{{p}}}
c^{}_{{n,m}_{{n}}}
\label{eq.10}
\end{eqnarray}
and 
\begin{equation}
[ {\hat{A}}^{s.p.}_k(t/2), {\hat{A}}^{s.p.}_k(-t/2) ] = 0.
\label{eq.11}
\end{equation}
It means that if we consider only the single particle part
of the nuclear Hamiltonian, the nuclear matrix $M_{GT}$ is just equal 
to zero. It is however expected since the $2\beta 2\nu$-decay  is a second 
order process correlated by the residual interaction. 
Without the residual interaction only the two-subsequent 
beta decay processes are possible, if they are energetically
allowed. Clearly, without the residual interaction it is not
possible to construct the spectrum of  the intermediate nucleus. 

We note that $M^{s.p.}_{{GT}}=0$ comes as a result of the cancelation 
between the direct and cross term of $M_{GT}$ in Eq.\ (\ref{eq.8}).
If we use $E_n-E_i+p_{20}+k_{20} = E_n-E_f-p_{10}-k_{10}$
and transform $M_{GT}$ in Eq.\ (\ref{eq.8}) in the integral 
representation we obtain  
\begin{eqnarray}
M_{{GT}}&=&\lim_{\epsilon\to 0} 
~i\int_{0}^{\infty}
<0^{{+}}_{{f}}|
{\hat{A}}_{{k}}(0) {\hat{A}}_{{k}}(-t)
{{e}}^{{{-it(}}p_{10}+k_{10}{{-}}\epsilon{{)}}  }
+\nonumber \\
&& ~~~~~~~~~~~~{\hat{A}}_{{k}}(t) {\hat{A}}_{{k}}(0) 
{{e}}^{{{-it(}}-p_{10}-k_{10}{{-}}\epsilon{{)}}  }
|0^{{+}}_{{i}}> {d}t.
\label{eq.12}
\end{eqnarray}
If we suppose ${\hat{A}}^{}_k(t) \approx {\hat{A}}^{s.p.}_k(t)$, 
integrate over time variable in Eq.\ (\ref{eq.12}) and use the
anticommutation relation of the particle operator we find a complete
cancelation between  both terms in the r.h.s. of Eq.\ (\ref{eq.12}).
It shows that the single particle operator of the nuclear Hamiltonian
plays a less important role in the evaluation of $2\beta 2\nu$-decay matrix
elements. This situation has not been noticed in Ref.\cite{27}
 in which
the approximation $\hat{H} \approx {\hat{H}}^{s.p.}$ was also
discussed and therefore the authors there came to a different conclusion.

\subsection{ The QRPA and RQRPA Hamiltonians}

The INA approach for the calculation of the $M_{GT}$ in Eq.\ (\ref{eq.8})
consists of  two QRPA diagonalizations related to the initial and
final nuclei. The corresponding initial and final QRPA Hamiltonians 
$\hat{H}^i$ and $\hat{H}^f$ take the forms 
\begin{eqnarray}
\hat{H}^i &=& const_i + \sum_{m_i J M} \Omega^{m_i} Q^{+ m_i}_{JM}Q^{m_i}_{JM},
\nonumber \\
\hat{H}^f &=& const_f + \sum_{m J M} \Omega^{m_f} Q^{+ m_f}_{JM}Q^{m_f}_{JM},
\label{eq.14}
\end{eqnarray}
which are connected with two sets of  intermediate nuclear states
\begin{eqnarray}
|m_i J M, i> = Q^{+ m_i}_{JM}|0_{qrpa}^+ >_{i} ~~~~~
|m_f J M, f> = Q^{+ m_f}_{JM}|0_{qrpa}^+ >_{f} 
\label{eq.15}
\end{eqnarray}
generated from initial and final nuclei, respectively.
Henceforth we use label "i" for initial and "f" for the final nuclei.
$\Omega^{m_{i,f}}$ is the energy of the
m-th intermediate state and the phonon creation operators 
 $Q^{+ m_{i,f}}_{JM}$ is defined as followed:
\begin{eqnarray}
Q^{+ m_i}_{JM}&=&\sum_{p n}( X^{m_i}_{(p n)J} A^+(pnJM)-
Y^{m_i}_{(p n)J} {\tilde{A}}(pnJM)), \nonumber \\
Q^{+ m_f}_{JM}&=&\sum_{p n}( X^{m_f}_{(p n)J} B^+(pnJM)-
Y^{m_f}_{(p n)J} {\tilde{B}}(pnJM)).
\label{eq.15a}
\end{eqnarray}
$X^{m_{i,f}}_{(p n)J}$ and $Y^{m_{i,f}}_{(p n)J}$ are forwards- and 
backwards- variational amplitudes, respectively. $A^+(pnJM)$
and $B^+(pnJM)$ are respectively boson creation operators of initial 
and final nuclei.

The time dependent axial current $A_k(t)$ can be 
expanded in the QRPA phonon basis as follows\cite{27}:
\begin{eqnarray}
{\hat{A}}^{QRPA}_k(t) &=&\sum_{pn m} <p||\sigma ||n>[ 
(u_pv_n X^m_{(pn)1}+v_p u_n Y^m_{(pn)1})e^{i\Omega_m t} Q^{+ m}_{1k} +
\nonumber \\
&&\phantom{\sum_{pn JM} <p||\sigma ||n> }
(v_p u_n X^m_{(pn)1}+u_pv_n Y^m_{(pn)1})e^{-i\Omega_m t} 
{\tilde{Q}}^{m}_{1k}].
\label{eq.16}
\end{eqnarray}
Here, $u$ and $v$ are the BCS occupation amplitudes. 

The expression in Eq.\ (\ref{eq.16}) allows us to calculate the
transition operator of the time dependent nuclear matrix element 
$M_{AA}(t)$ in Eq.\ (\ref{eq.6}). If we suppose the equivalence
between both QRPA Hamiltonians in (\ref{eq.14}) we obtain
\begin{equation}
[ {\hat{A}}^{QRPA}_k(t/2), {\hat{A}}^{QRPA}_k(-t/2) ] = const,
\label{eq.17}
\end{equation}
which implies that the $2\beta 2\nu$-decay  nuclear matrix element 
is equal to zero because the $2\beta 2\nu$-decay  operator should be at least
a two-body operator changing two neutrons into two protons. 
A generalization of the above discussion
to the RQRPA approach is straightforward and leads to the same conclusion.
It means that the suppression of the $2\beta 2\nu$-decay  nuclear 
matrix element is connected with the fact that it is a higher order effect
in the boson or renormalized boson expansion
of the nuclear Hamiltonian. 
We can obtain non-zero results only if we go beyond the first order
boson or renormalized boson Hamiltonians.

One can ask why non-zero results are obtained in the INA QRPA and
RQRPA calculations of $M_{GT}$. We believe that the following 
reasons could be the origin of this problem:
i) The initial and final QRPA ground states are not orthogonal. 
Therefore, even for a constant transition operator non-zero 
results could be obtained.
ii) The particle number non-conservation. We note that even the 
average particle numbers of protons and neutrons for the excited 
states of the intermediate nucleus differ from the correct ones. 
iii) There is a mismatching between the initial and final QRPA Hamiltonians
and as a consequence the two sets of intermediate nuclear
states generated from initial and final nuclei are not orthogonal 
to each other.  In the QRPA  or  RQRPA 
calculation of the $M_{GT}$ one arrives at the formula:
\begin{equation}
M^{2\nu}_{GT}=\sum_{m^{}_{i},m^{}_{f} k}
\frac{^{}_{f}<0_{qrpa}^+|{\tilde{\hat{A}_k(0)}}|1_{m^{}_{f}}^+>
<1_{m^{}_{f}}^+ | 1_{m^{}_{i}}^+> 
<1_{m^{}_{i}}^+ |\hat{A}_k(0)|0_{qrpa}^+>_{i}^{}}
{\Omega^{m^{}_{f}}_{1^+}+\Omega^{m^{}_{i}}_{1^+}}
\label{eq.18}  
\end{equation}
Here, $<1_{m^{}_{f}}^+ | 1_{m^{}_{i}}^+>$ is the overlap factor 
of the intermediate nuclear states generated from initial and final 
nuclei, given by\cite{2}
\begin{equation}
<1_{m^{}_{f}}^+ | 1_{m^{}_{i}}^+> =
\sum_{p n}~\big(
{X}^{m_{i}^{}}_{(p n)1^+}
{X}^{m_{f}^{}}_{(p n)1^+} -
{Y}^{m_{i}^{}}_{(p n)1^+}
{Y}^{m_{f}^{}}_{(p n)1^+} \big).
\label{eq.19}  
\end{equation}
This overlap factor has been considered practically in all QRPA 
or RQRPA calculations of $2\beta 2\nu$-decay process. 
In the case in which  the two sets of intermediate nuclear states deduced from
initial and final nuclei are identical  Eq.\ (\ref{eq.18}) is 
just the orthonormal condition for two QRPA states. However, this is  not  the
case in a realistic calculation and we shall show later that we 
can hardly expect it within the QBA. In addition we note that
the phases of the two sets of intermediate states are in principal arbitrary.
Therefore, it is necessary to identify them e.g. 
by requiring the diagonal elements of the overlap matrix to be positive
or by requiring the largest component of the wave function for each
state to be positive. 

The equivalence of the two sets of the intermediate nuclear states 
is connected with the equivalence of both QRPA Hamiltonians in 
Eq.\ (\ref{eq.14}). Let discuss this point within the QBA.

The quasiparticle creation and annihilation operators of the
initial  ($a^+, a$) and final ($b^+, b$) nuclei are connected with 
the particle creation and annihilation ($c^+, c$) operators by the 
BCS-transformations. As a consequence
there is a unitary transformation between the initial 
and final quasiparticles both for protons and neutrons. 
In the case of proton quasiparticles it takes the form:
\begin{eqnarray} 
\left( \matrix{ a^+_p  \cr 
a^{}_{\tilde{p} }} \right) &= &
\left( \matrix{ u^i_p & -v^i_p  \cr v^i_p & ~~u^i_p} \right) 
\left( \matrix{ u^f_p & ~~v^f_p  \cr -v^f_p & u^f_p} \right) 
\left( \matrix{ b^+_p \cr 
 b^{}_{\tilde{p}} } \right),\nonumber \\
&&
\left( \matrix{ {\tilde{u}}_p & -{\tilde{v}}_p  \cr 
{\tilde{v}}_p & ~~{\tilde{u}}_p} \right) 
\left( \matrix{ b^+_p \cr 
 b^{}_{\tilde{p}} } \right),  
\label{eq.20}  
\end{eqnarray} 
Relation  (\ref{eq.20})
allows us, by using the QBA, to 
rewrite the boson operators of the initial nucleus with the help of the boson 
operators of the final nucleus:
\begin{equation}
A^+(pn JM) = {\tilde{u}}_p{\tilde{u}}_n B^+(pn JM) -
{\tilde{v}}_p{\tilde{v}}_n {\tilde{B}}(pn JM) 
\label{eq.21}  
\end{equation}
We note that $|\tilde{u}| \approx 1$ and $|\tilde{v}| \approx 0$.
Therefore, we shall omit further terms proportional to $\tilde{v} \tilde{v}$.
Next, we can rewrite $Q^{+ m_i}_{JM}$  with the 
$Q^{+ m_f}_{JM}$ and ${\tilde{Q}}^{ m_f}_{JM}$ as follows:
\begin{equation}
Q^{+ m_i}_{JM} = \sum_{m_f} \big( {\cal{O}}^{m_i m_f}_J Q^{+ m_f}_{JM} +
{\cal{P}}^{m_i m_f}_J {\tilde{Q}}^{ m_f}_{JM} \big),
\label{eq.22}  
\end{equation}
where,
\begin{equation}
{\cal{O}}^{m_i m_f}_J =
\sum_{p n}~{\tilde{u}}_p {\tilde{u}}_n \big(
{X}^{m_{i}^{}}_{(p n)J}
{X}^{m_{f}^{}}_{(p n)J} -
{Y}^{m_{i}^{}}_{(p n)J}
{Y}^{m_{f}^{}}_{(p n)J} \big),
\label{eq.23}  
\end{equation}
\begin{equation}
{\cal{P}}^{m_i m_f}_J =
\sum_{p n}~{\tilde{u}}_p {\tilde{u}}_n \big(
{X}^{m_{i}^{}}_{(p n)J}
{Y}^{m_{f}^{}}_{(p n)J} -
{Y}^{m_{i}^{}}_{(p n)J}
{X}^{m_{f}^{}}_{(p n)J} \big). 
\label{eq.24}  
\end{equation}
It is now straightforward  to rewrite $H^i$ with the phonon operators 
of the final nucleus and to perform a comparison with the $H^f$.
The equivalence of both Hamiltonians requires that the following
relations are fullfiled:
\begin{equation}
\sum_{m_i} \Omega^{m_i}_J \big( 
{\cal{O}}_J^{m_i m_f} {\cal{O}}_J^{m_i {m'}_f} +
{\cal{P}}_J^{m_i m_f} {\cal{P}}_J^{m_i {m'}_f}\big) = 
\Omega^{m_f}_J \delta_{m_f {m'}_f}
\label{eq.25}  
\end{equation}
\begin{equation}
\sum_{m_i} \Omega^{m_i}_J \big( 
{\cal{O}}_J^{m_i m_f} {\cal{P}}_J^{m_i {m'}_f} \big) = 0.
\label{eq.26}  
\end{equation}

We know that in the vicinity of the collapse of the QRPA solution
the lowest $1^+$ state of the
intermediate nucleus plays an important role in the calculation 
of $M_{GT}$ as
it is strongly influenced by the ground state correlations.
Therefore, there is an
interest to check the validity of the expression (\ref{eq.25}) for
this state. 
We note that the above expressions could be used also in the case of
the renormalized QBA scheme,
if we replace the $X$ and $Y$ amplitudes with 
the  renormalized amplitudes $\overline{X}$, $\overline{Y}$
(see Ref.\cite{10})  and suppose for the 
renormalized factors the following relation    
${\cal D}^i_{(p n)1^+ }/{\cal D}^f_{(p n)1^+ } \simeq 1$. 
The lowest energy of the intermediate state $\Omega^{m_f}_{1^+}$
for $2\beta 2\nu$-decay   of 
$^{76}Ge$ obtained directly from the final nucleus 
by the QRPA and RQRPA diagonalization and 
indirectly with help of expression (\ref{eq.25}) 
for three different model spaces given in Ref.\cite{28}
is presented in Fig.~\ref{fig.1}.
By glancing at Fig. ~\ref{fig.1}. we see that close to the collapse of the
pn-QRPA the equivalence of the initial and final 
QRPA Hamiltonians  is getting worse with increasing
parameter $g_{pp}$. It is more apparent if a larger model space is used. 
We remark  that the accuracy of the pn-RQRPA calculation of the 
lowest $1^+$ state for a physically acceptable value of $g_{pp}$
is considerably better in respect to the QRPA one.
It clearly shows that the RQRPA offers a more reliable solution. 
However, the initial and final RQRPA Hamiltonians still remain different
and this could be one of the reasons of the  non-zero results 
obtained in the INA calculation of $M_{GT}$.

For the sake of completeness 
we also plot in  Fig. ~\ref{fig.2}
the $M_{GT}$ of the $2\beta 2\nu$-decay   
of  $^{76}Ge$ calculated within the INA pn-QRPA and 
pn-RQRPA approach for three different model spaces.
We see that in both cases  $M_{GT}$ demonstrates an increased
sensitivity to $g_{pp}$ with enhanced model
space. 
We note also that the pn-RQRPA allows to perform calculations 
far behind the collapse of the pn-QRPA. These calculations 
have been performed with an overlap factor of the initial
and final states, which has been approximated as follows:
\begin{equation}
<1_{m^{}_{f}}^+ | 1_{m^{}_{i}}^+> \approx 
[Q^{ m_f}_{JM}, Q^{+ m_i}_{JM}] =
 {\cal{O}}_J^{m_f m_i} 
\label{eq.27}  
\end{equation}
The advantage of this overlap factor over the  one of 
Eq.\ (\ref{eq.19}) is that now the results are independent of the
phases of the quasiparticle states, which are in principal 
arbitrary [For a given quasiparticle eigenenergy E, there are two
solutions (u,v) and (-u,-v)]. 
For the  overlap factor of Eq.\ (\ref{eq.19})
this condition is not fullfiled. This can be easily 
proved by a numerical test. 

It is also of interest to evaluate $M_{GT}$ by using the overlap
functions ${\cal{O}}_1^{m_i m_f}$ and ${\cal{P}}_1^{m_i m_f}$
and only one of the two RQRPA Hamiltonian, e.g. the ${\hat{H}}^i$.
With the help of Eqs.\ (\ref{eq.16}) and (\ref{eq.22}) one has for 
the $\beta^+$ transition 
amplitudes:
\begin{equation}
<1_{m^{}_{i}}^+ ||\hat{A}(0)||0_{qrpa}^+>_{i}^{} = 
\frac{1}{\sqrt{3}}\sum_{p n }\sqrt{D^i_{(pn)1}}
[ u^i_p v^i_n \overline{X}^{m_i}_{(p n)1}
+v^i_p u^i_n \overline{Y}^{m_i}_{(p n)1}]
\label{eq.27a}  
\end{equation}
\begin{eqnarray}
^{}_{f}<0_{qrpa}^+||{\widetilde{\hat{A}(0)}}||1_{m^{}_{f}}^+> = 
\frac{1}{\sqrt{3}}\sum_{p n m_i}
[ (v^i_p u^i_n \overline{X}^{m_i}_{(p n)1}+
u^i_p v^i_n \overline{Y}^{m_i}_{(p n)1})
{\cal{O}}_1^{m_i m_f} \nonumber \\
 +
(u^i_p v^i_n \overline{X}^{m_i}_{(p n)1}+
v^i_p u^i_n \overline{Y}^{m_i}_{(p n)1})
{\cal{P}}_1^{m_i m_f}]\sqrt{D^i_{(pn)1}}.
\label{eq.27b}  
\end{eqnarray}
The $M_{GT}$ calculated with help of only one pn-RQRPA Hamiltonian 
is drawn in Fig.~\ref{fig.3}. We note a large discrepancy between  
the results obtained with initial and final pn-RQRPA Hamiltonians and those 
obtained by the standard INA calculation. In this way one sees that 
the $2\beta 2\nu$-decay matrix elements are very sensitive to the 
restrictions of the RQBA approximation. 
From the Fig.~\ref{fig.3} it follows also that the INA calculation has 
been mostly influenced  by the final pn-RQRPA Hamiltonian. It is  an
indication that the main problem of the INA consist in the non-orthogonality
of the initial and final ground states.

The main message of this  Section is  that the $2\beta 2\nu$-decay 
is a higher order effect in the boson expansion of the nuclear 
Hamiltonian. Several procedures have been proposed
which outline the importance of higher order effects and
try to take into account  
some higher-order QRPA corrections\cite{5}. 
Maybe, the most perspective way is the fermion-boson 
mapping procedure discussed by M. Sambataro, F. Catara and J. Suhonen
\cite{29} within a schematic model. In this way it is  expected  that
the correspondence between the initial and final nuclear Hamiltonians 
will be improved and more reliable results could be obtained. 

It is worthwhile to notice that the usual strategy has been 
first to try to reproduce 
the observed $2\beta 2\nu$ - decay half times  within a given 
nuclear model in order to gain confidence in the calculated 
$2\beta 0\nu$ - decay nuclear matrix elements. However, 
there is a principal difference from the nuclear physics point of view. 
The $0\beta 2\nu$ - decay is not a higher order effect in the boson
expansion of the nuclear Hamiltonian and therefore the  renormalized QBA
scheme could be sufficient.
It is because the nucleons undergoing beta decays in the nucleus
are correlated by the exchange of Majorana neutrinos  
and then the decomposition in Eq.\ (\ref{eq.3}) is irrelevant. 
We remind the reader
 that the QRPA and RQRPA Hamiltonians have been found successful
in describing the single beta decay transition. 

\subsection{ The Operator Expansion Method}

The OEM is a nuclear structure method for the $2\beta 2\nu$-decay,
which has the advantage to avoid the explicit sum over the intermediate
nuclear states. There are two different ways to derive 
the $2\beta 2\nu$-decay OEM transition operators. 
In an approach proposed by Ching and Ho (OEM1)\cite{14} the expansion of the
denominators of $M_{GT}$ in Eq.\ (\ref{eq.12}) in  Taylor series
is used. In a different approach proposed by \v Simkovic (OEM2)\cite{15}
the OEM is derived from the integral representation of the nuclear
matrix element $M_{GT}$ in Eqs.\ (\ref{eq.5}),
(\ref{eq.6}) and (\ref{eq.7}). 
It has been found that the OEM2 offers advantages over OEM1 as there 
are no problems of convergence  in the power series expansion
of the denominator, which has been a subject of criticism,  
J. Engel et al.\cite{30}. 

In the OEM the $2\beta 2\nu$-decay transition operator is described by an 
infinite series of multiple commutators of the nuclear Hamiltonian H
and the Gamow-teller transition operator A(0) [see Eq.\ (\ref{eq.6})
and (\ref{eq.7})]. 
The OEM is based on two main assumptions\cite{14,15}: 
i) It is assumed that the
kinetic energy operator can be ignored in the resulting commutators 
and therefore  the 
nuclear Hamiltonian is represented only by the two-body interaction 
terms. 
ii) Only two-body terms are retained in  evaluating each commutator
and higher order terms are neglected. 
These two  approximations seem reasonable. 
The omission of the one-body terms of the nuclear Hamiltonian 
 is justified by the fact that these terms play a less important role in the
evaluation of $M_{GT}$. It is easy to see it  if we 
consider only the one-body  part of the nuclear Hamiltonian $H_0$,
then $A(t) = e^{iH_0t}A(0)e^{-iH_0t}$ as well as 
the commutator [A(t/2), A(-t/2)] are one body operators. 
However, there are no contributions  to $2\nu 2\beta$-decay 
from a one body operator, as  the $2\nu 2\beta$-decay 
operator should be at least a two-body operator changing two 
neutrons into two protons. In respect to the second approximation
from the discussion in Section 3 it follows that this approximation 
goes beyond the QBA or renormalized QBA. In the case of the QBA 
the commutator [A(t/2), A(-t/2)] is just a constant but within the
OEM the commutator 
[A(t/2), A(-t/2)] is a two-body transition operator changing 
two neutrons into two protons. So, it is not  true 
that the OEM is an approximation to the QRPA as
it was believed before\cite{20}. 

Recently, the OEM has been reconsidered\cite{23}. 
It has been shown that the Coulomb interaction plays a 
decisive role within the OEM. 
Gmitro and \v Simkovic\cite{16} were first to introduce a
Coulomb interaction term 
\begin{eqnarray}
V_C&=&\frac{1}{2}\sum_{i\neq j} ~g_c(r_{ij}) ~O^\tau_{ij},
~~~~~~O^\tau_{ij}=\frac{1}{4}(1+\tau^0_i )(1+\tau^0_j ),
\label{eq.28} 
\end{eqnarray}
in the OEM formalism. Later it was done by Muto\cite{20}, who however obtained
different OEM transition operators. We shall discuss this important point
and explain the origin of these differences. 

The time dependent nuclear matrix element $M_{AA}(t)$ in (\ref{eq.7}) 
can be transformed to the following form:
\begin{eqnarray}
M_{AA}(t) = e^{i ( E_i - E_f )\alpha }
<0^{{+}}_{{f}}|e^{i H \alpha }
\frac{1}{2} [ A_{{k}}(t/2), A_{{k}}(-t/2) ] 
e^{- i H \alpha } |0^{{+}}_{{i}}>.
\label{eq.29} 
\end{eqnarray}
Obviously $M_{AA}(t)$ does not depend on $\alpha$, if 
the nuclear states and their corresponding energies can be 
considered as the eigenstates and eigenvalues 
of the nuclear Hamiltonian H.
\begin{eqnarray}
H|0^+_{i}>=E_{{i}}|0^+_{{i}}>,
~~ H|0_{{f}}>=E_{{f}}|0^+_{{f}}>
\label{eq.30}
\end{eqnarray}

Using the machinery of the OEM
for the operator inside the brackets 
in the r.h.s. in Eq.\ (\ref{eq.29}) one obtains
\begin{eqnarray}
M_{AA}(t) &=& e^{i ( E_i - E_f )\alpha }
<0^+_f|{\cal P} \sum \limits_{n \ne m} 
e^{i g_c(r_{nm}) O^\tau_{nm} \alpha} \times  \nonumber \\
&&\tau^+_n \tau^+_m {\cal{V}}^{OEM}(t, r_{nm}, \Pi^\sigma_s(n,m), 
\Pi^\sigma_t(n,m), S_{nm}) 
|0^+_i>.
\label{eq.31} 
\end{eqnarray}
Here, $\Pi^\sigma_s$ and $\Pi^\sigma_t$ are projectors onto spin singlet
and triplet states and ${\cal V}^{OEM}$ is the two-body OEM potential.
${\cal{P}}$ denotes the Principle value integration.
It is apparent that $M_{AA}(t)$ is $\alpha $ dependent for 
 $\alpha \ne 0$.  It means that if the
nuclear Hamiltonian does not contain a Coulomb 
term\cite{14,15}$^{,}$\cite{17}$^{-}$\cite{19} 
 or contains  a Coulomb interaction in the form of 
Eq.\ ({\ref{eq.8})\cite{16,20}$^{,}$\cite{21},
the derivation of the OEM transition operator is inconsistent.   
For $\alpha = t/2$ we obtain the OEM transition operators of Muto\cite{20} 
and the formulae of Gmitro and \v Simkovic\cite{16} could be reproduced 
for $\alpha = 0$, which differ from each other.  

It is worthwhile to notice that the above results point out some more important
aspects. The analysis is clearly showing that the mass
difference  of the initial and final nuclei $E_i - E_f$
plays an important role in the calculation of $2\nu 2\beta$-decay. 

The above mentioned inconsistency of the OEM could be avoided if one considers
an effective Coulomb interaction term $V_C$\cite{22} 
\begin{eqnarray}
V_C&=&\frac{1}{2}\sum_{i\neq j} ~(E_f -E_i ) ~O^\tau_{ij}.
\label{eq.32} 
\end{eqnarray}
In this way the one-body terms of the nuclear Hamiltonian 
are not totally neglected. 

Let's consider the approximated two-body nuclear Hamiltonian H
containing central $V_{CN}$ and tensor $V_{TN}$ interactions 
in addition to the effective Coulomb interaction 
(the notation of Ref.\cite{20} is used):
\begin{eqnarray}
H \approx V_C + V_{CN} +V_{TN},
\label{eq.33} 
\end{eqnarray}
where
\begin{eqnarray}
V_{CN}&=&\frac{1}{2}\sum_{i\neq j} 
~[~~   (~ g_{SE}(r_{ij})~\Pi^r_e(ij) ~+~
      g_{SO}(r_{ij})~\Pi^r_o(ij)~ )~\Pi^\sigma_s(ij) +
\nonumber \\
&&\phantom{~~~\frac{1}{2}\sum_{i\neq j}}
    ( ~g_{TE}(r_{ij})~\Pi^r_e(ij) ~+~
      g_{TO}(r_{ij})~\Pi^r_o(ij)~ ) ~\Pi^\sigma_t(ij) ~~],
\\
V_{TN}&=&\frac{1}{2}\sum_{i\neq j} 
~(~ g_{TNE}(r_{ij})~\Pi^r_e(ij) + 
g_{TNO}(r_{ij} )~\Pi^r_o(ij)~)~S_{ij}.
\label{eq.34} 
\end{eqnarray}
Then, within the OEM approximations the infinite series of the
commutators in Eq.\ (\ref{eq.7}) could be summed using  the
formulae\cite{15,16}
\begin{eqnarray}
e^{i g P t} A e^{-i g P t} =
\frac{1}{2}( A+PAP+cos(2gt)(A-PAP)+isin(2gt)[P,A]),
\label{eq.35} \\
e^{i g O t} A e^{-i g O t} =
A-[O,[O,A]] + cos(gt)[O,[O,A]] +isin(gt)[O,A], 
\label{eq.36} 
\end{eqnarray}
for $~~P^2=1$ and $O^2=O$.
Then by performing the integration in Eq.\ (\ref{eq.5})
over t we obtain the nuclear matrix element $M_{GT}$ :
\begin{eqnarray}
M_{GT}&=&<0^+_f|\frac{1}{2} {\cal{P}}
\sum \limits_{i \ne j} \tau^+_i \tau^+_j
~(~{\cal V}^{singlet}(r_{ij})~\Pi^\sigma_s(ij) ~+ \nonumber \\
&&\phantom{ \sum \limits_{i \ne j} \tau^+_i \tau^+_j }
 {\cal V}^{triplet}(r_{ij})~\Pi^\sigma_t(ij) ~+~
 {\cal V}^{tensor}(r_{ij})~S_{ij}~)~|0^+_i>,
\label{eq.37} 
\end{eqnarray}
where, 
\begin{eqnarray}
{\cal V}^{singlet}&=&\frac{-2}{g_{TE}-g_{SE}-4g_{TNE}+\Delta } -
 \frac{4}{g_{TE}-g_{SE}+2g_{TNE}+ \Delta } \nonumber \\ 
{\cal V}^{triplet}&=&\frac{1}{3} [ 
\frac{4}{\Delta} +\frac{4}{-6g_{TNO}+\Delta}+
\frac{4}{6g_{TNO}+\Delta}  -  \nonumber \\
&& \frac{2}{g_{SO}-g_{TO}+4g_{TNO}+\Delta } -
\frac{4}{g_{SO}-g_{TO}-2g_{TNO}+\Delta } ], \nonumber \\ 
{\cal V}^{tensor}&=& \frac{1}{3} [ \frac{1}{\Delta}+
\frac{1}{-6g_{TNO}+\Delta } -
\frac{2}{6g_{TNO}+\Delta } + \nonumber \\
&&\frac{1}{g_{SO}-g_{TO}+4g_{TNO}+\Delta } -
\frac{1}{g_{SO}-g_{TO}-2g_{TNO}+\Delta } ].
\label{eq.38} 
\end{eqnarray}
If we neglect the tensor potential $V_{TN}$ we have
\begin{equation}
M_{GT}=<0^+_f|\frac{1}{2} {\cal{P}}
\sum \limits_{n \ne m} \tau^+_n \tau^+_m
(f^0(r)\Omega^0_{nm} + f^1(r)\Omega^1_{nm})|0^+_i>
\label{eq.39} 
\end{equation}
\begin{eqnarray}
f^0(r)&=&  \frac{-6}{g_{TE}(r)-g_{SE}(r)+\Delta} \nonumber \\
f^1(r)&=&   \frac{4}{\Delta }-
\frac{2}{g_{SO}(r)-g_{TO}(r)+\Delta }
\label{eq.40} 
\end{eqnarray}
If we neglect both central and tensor potential $V_{TN}$ 
we obtain
\begin{equation}
M_{GT}=<0^+_f|\frac{1}{2} 
\sum \limits_{n \ne m} \tau^+_n \tau^+_m
\frac{2}{\Delta}\vec{\sigma}_n\cdot \vec{\sigma}_m |0^+_i>.
\label{eq.41} 
\end{equation}
We note that the OEM transition operators obtained by  Muto\cite{20} differ 
from the above ones.This is due to the inconsistent derivation
in the framework of the OEM1. The approximate Hamiltonian
considered by Muto does not allow to fix the difference of the masses of
the initial
and final nuclei, which is a basic feature of the $2\beta 2\nu$-decay
matrix element. Thus, one obtains different transition operators for 
different values of the parameter $\alpha$  in Eq.\ (\ref{eq.31}).
This important result is coming from the Heisenberg nature of the
axial current operators and it can been hardly  seen 
within the OEM1 method proposed by Ching and Ho\cite{14}. 
The importance of the mass difference  between the initial and final
nuclei becomes apparent within the OEM2\cite{22}, which contains
elements of the field theory. In addition, the OEM2 shows explicitly that 
only the Principal value part of the  ${\cal V}^{OEM}$ potential is relevant
for the nuclear matrix element $M_{GT}$. A fact which 
is not clear within the OEM1.
We note that the poles of the ${\cal V}^{OEM}$ potential appear for 
$r_{nm} \approx 1.5-2.0$ fm. It is because of the small energy release
for this process ($\Delta \approx 2$ MeV). It means that only the 
long range part of the nucleon-nucleon interaction plays an important
role in the calculation and that the ${\cal V}^{OEM}$  potential
is expected to be independent of the chosen type of the nucleon-nucleon
interaction. It is worthwhile to notice
that the consistent OEM-potential in Eqs.\ (\ref{eq.37})-(\ref{eq.41})
does not vanish even for a zero-range $\delta$-force. 

For the calculation of $M_{GT}$ in Eq.\ (\ref{eq.36}) it is
necessary to know the wave functions of the initial and final nucleus.
The OEM could be combined with the ground state wave functions 
of the QRPA or renormalized RQRPA models\cite{17}$^{-}$\cite{21} (OEM+RPA)
\begin{eqnarray}
M_{GT}^{OEM+RPA} =
\sum_{{p n \acute{p} \acute{n} } \atop{ J^{\pi}
m_i m_f {\cal J}  }}
~(-)^{j_{l}+j_{k'}+J+{\cal J}}(2{\cal J}+1)
\left\{
\matrix{
j_p &j_n &J \cr
j_{n'}&j_{p'}&{\cal J}}
\right\}\times  
\nonumber \\
{^{}_{f}<}0_{rpa}^+\parallel 
\widetilde{[c^+_{p'}{\tilde{c}}_{n'}]_J}\parallel J^\pi m_f>
<J^\pi m_f|J^\pi m_i>
<J^\pi m_i \parallel [c^+_{p}{\tilde{c}}_{n}]_J \parallel 0^+_{rpa}>_i 
\nonumber \\
\times <p,p';{\cal J}|{\cal{P}}
\tau_1^+ \tau_2^+ {\cal{V}}^{OEM}(t, r_{12}, \Pi^\sigma_s(1,2), 
\Pi^\sigma_t(1,2), S_{1,2}) 
|n,n';{\cal J}>~~~~~~
\label{eq.42}   
\end{eqnarray}
In the previous OEM+QRPA calculations with the inconsistent
OEM potential ${\cal{V}}^{OEM}$,
 the $M_{GT}^{OEM+RPA}$ has been found not sensitive
to  the strength of the particle-particle interaction\cite{17}$^{-}$\cite{20}. 
Recently,
the $M_{GT}^{OEM+RPA}$  has been calculated 
for the $2\beta 2\nu$-decay of $^{76}Ge$ within the
renormalized pn-QRPA with the
overlap factor in Eq.\ (\ref{eq.19}) and a  consistent
OEM-potential in Eqs.\ (\ref{eq.38})-(\ref{eq.38})
\cite{23}. 
Small and large model spaces 
containing respectively the full $2-4\hbar\omega$ the full $0-5\hbar\omega$ 
major oscillator shells have been considered.
A strong suppression of the results with increasing 
$g_{pp}$ has been found\cite{23}.
The sensitivity of $M_{GT}^{OEM+RPA}$ to $g_{pp}$ within the physically
acceptable region of particle-particle strength has been increased
considerably with the enhancement of the model space. It is clear that
this effect could have its origin only in the pn-RQRPA wave functions. 
We recall that the OEM-potential is a two-body operator, which represents
a sum over all intermediate nuclear states and it is independent of the basis.
Then, the instability of the results has to be related to the
overlap factor and to the fact that the initial and
final states are not orthogonal. 

We mentioned already that
the  overlap factor of Eq.\ (\ref{eq.19})
does not guarantee the independence of the
results in respect to the phases of the BCS states, which are 
arbitrary.  We have found that the $2\beta 2\nu$-decay matrix element 
is very sensitive to this problem. One can see it by calculating
$M_{GT}^{OEM+RPA}$ with different overlap factors, i.e. a
delta function overlap factor and the overlap factors  of Eqs. (\ref{eq.19}) 
and (\ref{eq.27}). The uncertainty of the results 
is build in into the calculation through the pn-RQRPA wave functions.
It is worth to notice  that a similar problem appears by calculating 
the closure matrix element,
\begin{equation}
M_{clos}=<0^+_f|
\sum \limits_{n \ne m} \tau^+_n \tau^+_m
\vec{\sigma}_n\cdot \vec{\sigma}_m |0^+_i>.
\label{eq.43} 
\end{equation}
We note that $M_{clos}$ differs only by a constant with the 
$M_{OEM}$ in the case in which central and tensor interactions are neglected.
From the above discussion it follows that OEM should be not 
combined with QRPA or RQRPA. Nevertheless, there is an interest for 
OEM calculation with other ground state wave functions especially 
with shell model ones. 

\section{ Conclusions}

In summary, the integral representation of the nuclear matrix element
$M_{GT}$ in Eq.\ (\ref{eq.5}), (\ref{eq.6}) and (\ref{eq.7})
has been found useful for a critical analysis of different 
approximation schemes. The single-particle approximation of the
nuclear Hamiltonian implies $M_{GT}$  to be equal to zero because
of the mutual cancelation of the direct and cross terms. It means that
without the residual interaction there is no $2\beta 2\nu$-decay
and only the two subsequent beta decays
are possible, if they are energetically allowed. We note
that without the residual interaction no excited states could be generated. 

The use of the pn-QRPA and pn-RQRPA nuclear Hamiltonians reveals that 
the $2\beta 2\nu$-decay transition operator is a constant, if one
assumes that the initial and final nuclear Hamiltonians correspond
to each other. However, $2\beta 2\nu$-decay operator should be at least
a two-body operator changing two neutrons into two-protons.
Therefore, the inclusion of higher order terms of the boson expansion
of the nuclear Hamiltonian is necessary for a QRPA and RQRPA treatment of
the $2\beta 2\nu$-decay process. It is worth mentioning that there is 
no such requirement in the case of the single beta decay 
and $2\beta 0\nu$-decay calculations.

By using a unitary transformation between the quasiparticles of 
the initial and final uncorrelated BCS ground states 
the problem of the equivalence
of the initial and final pn-QRPA (pn-RQRPA) nuclear Hamiltonians
has been studied. It has been found that the pn-RQRPA Hamiltonians 
demonstrate  a better mutual agreement like the pn-QRPA ones.  
The mismatching between  both pn-QRPA Hamiltonians is large 
and their correspondence is getting worse with
increasing model space and  particle-particle strength parameter 
$g_{pp}$. 

Previously, it was believed that the OEM is an approximation to the
QRPA approach\cite{20}. However, the $2\beta 2\nu$-decay transition 
operator within the OEM is a two-body operator and not a
constant as within the QRPA or RQRPA.
It means that the OEM approximations
go beyond the QBA and RQBA. We note that
the previous derivation\cite{20} of the $2\beta 2\nu$-decay OEM transition
operator was inconsistent because the role of the energy difference
between the initial and final nuclear ground state in the calculation was
overlooked. We present the OEM potential derived in a consistent way
by considering the effective Coulomb interaction term. 
In this way the 
one-body terms of the nuclear Hamiltonian are not totally 
neglected. A combination of the OEM transition operator with the
pn-RQRPA wave functions reflect the instabilities incorporated
in the pn-RQRPA treatment of the two-vacua problem. Therefore,
we suppose that the combination of  the OEM-transition operator 
with other ground state way functions, e.g. shell model ones,
could be perhaps more predictive for the $2\beta 2\nu$-decay process. 

We remark that the OEM is a special method developed for the
$2\beta 2\nu$-decay, which could perhaps find a wider use for the
study of the pion p-wave contribution to the process of Double Charge Exchange
(DCX) pion on nuclei. The application of the OEM method is limited to 
the problems, where the nucleon-nucleon correlations by a meson
exchange plays a dominant role. However, this is not the case in 
single beta decay and  neutrinoless double beta decay calculations
(there is a neutrino correlation of two beta decays in the nucleus).
For these processes the one-body term of the nuclear Hamiltonian
is expected to play a crucial role. Therefore, 
methods constructing explicitly the intermediate nucleus spectrum 
could be more successful in the treatment of these processes.

The presented studies have shown that the two-vacua problem appearing
in the calculations of the $2\beta 2\nu$-decay matrix elements are not
safely treated within the QBA or RQBA. As a result there are two-sets 
of intermediate nuclear states generated from initial and final nuclei,
which do not correspond to each other. It is because of the 
particle number non-conservation and the violation of the Pauli
exclusion principle.  Both these problems are connected to each  
other because one can not guarantee the particle number conservation 
without an explicit consideration of the Pauli principle. We note that
the RQBA takes into account the Pauli principle only in a approximate
way. In the QRPA and RQRPA INA calculation of the 
$2\beta 0\nu$-decay matrix element the inaccuracy of the 
considered approximations for solving the two-vacua problem
is covered by the introduced overlap factor. In this paper
we have proposed a less critical way to evaluate it. We believe
that by including higher order terms of the boson expansion
of the nuclear Hamiltonian the equivalence of both nuclear Hamiltonians
and of the  two sets of the intermediate nuclear states  will
be improved and the overlap factor will be closer to a delta function.
A schematic study of M. Sambataro et al\cite{29}, 
based on the fermion-boson mapping procedure is rather encouraging 
and could lead to a better  agreement of the
two-different nuclear Hamiltonians and to  more reliable
results.

\begin{figure}[t]
\caption{The energy of the lowest $1^+$ state of the intermediate nucleus
$^{76}As$ calculated directly
within the pn-QRPA and pn-RQRPA for $^{76}Se$ isotope
and indirectly with the help 
of  Eq.\ (\ref{eq.25}) is plotted as 
function of the particle-particle coupling constant $g_{pp}$
for a 12-level (the full $2-4\hbar\omega$ major oscillator shells)
and a 21-level model space (the full $0-5\hbar\omega$ major 
oscillator shells).}
\label{fig.1}
\end{figure}

\begin{figure}[t]
\caption{The nuclear 
matrix element $M_{GT}$ for the $2\nu 2\beta$-decay of $^{76}Ge$ 
calculated within the pn-QRPA and pn-RQRPA INA, is plotted as  a
function of the particle-particle coupling constant $g_{pp}$.
The dashed line corresponds to the 9-level model space
(the full $3-4\hbar\omega$ major oscillator shells),
the dot-dashed line to the 12-level model space
(the full $2-4\hbar\omega$ major oscillator shells)
and the solid line to the 21-level model space
(the full $0-5\hbar\omega$ major oscillator shells).} 
\label{fig.2}
\end{figure}

\begin{figure}[t]
\caption{The calculated nuclear 
matrix element $M_{GT}$ for the $2\nu 2\beta$-decay of $^{76}Ge$ 
pn-RQRPA INA plotted as  a
function of the particle-particle coupling constant $g_{pp}$
for a 12-level and a 21-level model space.
The solid line corresponds to the standard calculation with the
initial and final QRPA Hamiltonians. The dashed line is the calculation
which considers only the initial QRPA Hamiltonian and the overlap functions 
[see Eqs.\ (\ref{eq.27a}) and (\ref{eq.27b})]
and the dot-dashed line the final QRPA 
Hamiltonian and the overlap functions.}
\label{fig.3}
\end{figure}

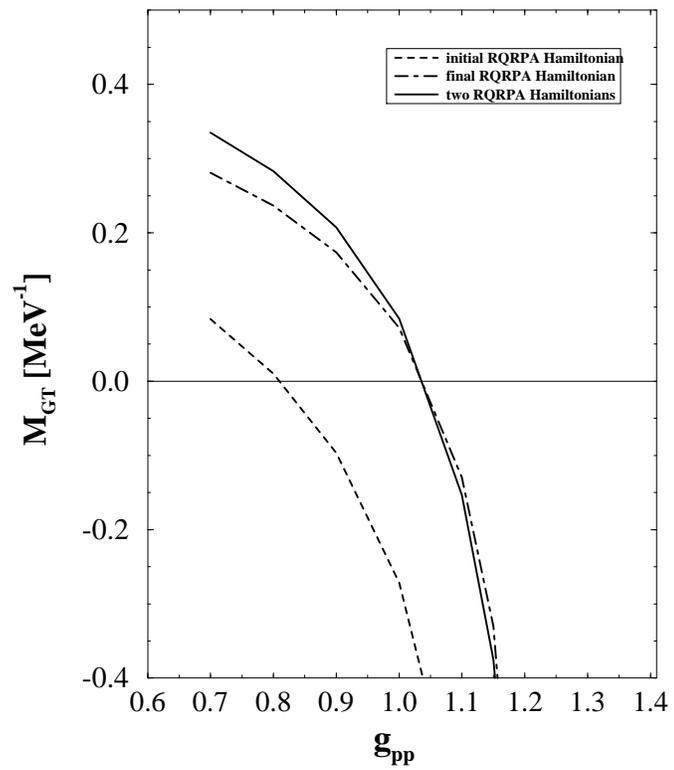
\end{document}